\documentclass[global,twocolumn]{svjour} 
\usepackage{latexsym}
\usepackage{graphics}
\journalname{
}
\begin{document}
\title{Surface-sensitive NMR in optically pumped semiconductors}
\author{Atsushi Goto\inst{1,2}
 \and Tadashi Shimizu\inst{1}
 \and Kenjiro Hashi\inst{1}
 \and Shinobu Ohki\inst{1}
}                     
%
%
\institute{National Institute for Materials Science, Tsukuba, Ibaraki 305-0003, Japan
\and PRESTO, Japan Science and Technology Agency, 4-1-8 Honcho, Kawaguchi, Saitama 332-0012, Japan.}
\date{
}
%
\maketitle
\begin{abstract}
We present a scheme of \textit{surface-sensitive 
nuclear magnetic resonance} in optically pumped semiconductors,
where an NMR signal from a part of the surface of a bulk compound semiconductor
is detected apart from the bulk signal.
It utilizes optically oriented nuclei with a long spin-lattice relaxation time
as a polarization reservoir for the second (target) nuclei to be detected.
It provides a basis for the \textit{nuclear spin polarizer}
[IEEE Trans. Appl. Supercond. \textbf{14,} 1635 (2004)],
which is a polarization reservoir at a surface of the optically pumped semiconductor
that polarizes nuclear spins in a target material
in contact through the nanostructured interfaces.
\end{abstract}

\section{Introduction}
\label{intro}

\textit{Spatial resolution} is one of the key ingredients
in nano-scale characterizations.
Nuclear magnetic resonance \\ (NMR),
which is a powerful analytical tool to investigate structural 
and dynamical properties of materials,
is less advantageous in this respect.
To gain high spatial resolution in NMR,
one needs to invoke some characteristic systems that are specially designed for this purpose.
Examples include magnetic resonance imaging microscopy\cite{ciobanu02}
and magnetic resonance force microscopy,\cite{ruger04}
where high spatial resolution is achieved by
a strong magnetic field gradient that surpasses NMR line widths.
This requirement, as well as that for high sensitivity 
to detect small number of nuclei in a small region,
has posed serious challenges to these techniques.

\begin{figure}[b]
\begin{center}
\resizebox{0.35\textwidth}{!}{%
  \includegraphics{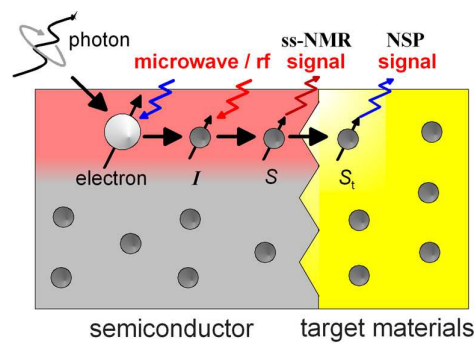}
}
\caption{
Successive polarization transfer process in a compound semiconductor
(with $I$ and $S$ nuclei) and a `target' material (with $S_t$).
Initial polarization is provided by circularly polarized photons,
which excite polarized electrons at a surface of the semiconductor. 
The electrons polarize $I$ nuclei nearby via hyperfine couplings,
and the $I$ nuclei polarize $S$ via heteronuclear couplings.
As a result, enhanced $S$ signals from the illuminated area are detected 
(\textit{surface-sensitive NMR}).
The polarization is further transferred to $S_t$ across the interfaces, 
and enhanced $S_t$ signals in the target material are detected 
(\textit{nuclear spin polarizer}).
In these processes, signals from the nuclei located deep in the bulk (gray spheres)
are cancelled out and do not contribute to the signal detected.
}
\label{spt}
\end{center}
\end{figure}
\begin{figure}[t]
\begin{center}
\resizebox{0.35\textwidth}{!}{%
  \includegraphics{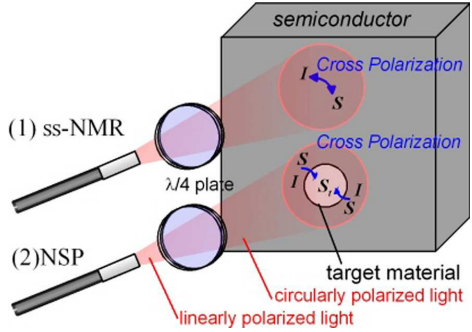}
}
\caption{
Schematic illustrations of (1) \textit{surface-sensitive NMR} (\textit{ss-NMR}) 
and (2) \textit{nuclear spin polarizer} (\textit{NSP}).
A linearly polarized light emitted from a fiber end
is converted to a circularly polarized one with a quarterwave plate,
and applied to the surface of a bulk compound semiconductor,
where $I$ polarization is created.
In the \textit{surface-sensitive NMR}, the polarization is transferred to $S$
inside the illuminated area by cross-polarization,
while in \textit{NSP}, 
it is transferred to $S_t$ in the target material at a surface of the semiconductor.}
\label{illumination}
\end{center}
\end{figure}

\begin{figure}[b]
\begin{center}
\resizebox{0.25\textwidth}{!}{%
  \includegraphics{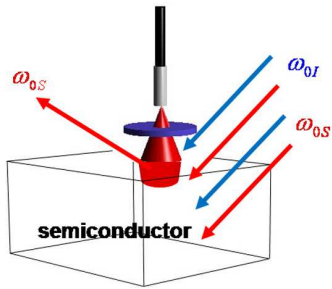}
}
\caption{
RF pulses irradiated and a signal retrieved in the \textit{surface-sensitive NMR}.
The rf pulses with the angular frequencies 
of $\omega_{0I}$ and $\omega_{0S}$ excite both the $I$ and $S$ spins in the whole sample.
On the other hand, the retrieved rf signal is only that with $\omega_{0S}$ 
from the spot illuminated with the light.
The (penetration) depth of the light is controlled by the wavelength 
near the band gap energy.\cite{michal99}}
\label{ss-nmr}
\end{center}
\end{figure}

\begin{figure}[b]
 \begin{center}
\resizebox{0.35\textwidth}{!}{%
  \includegraphics{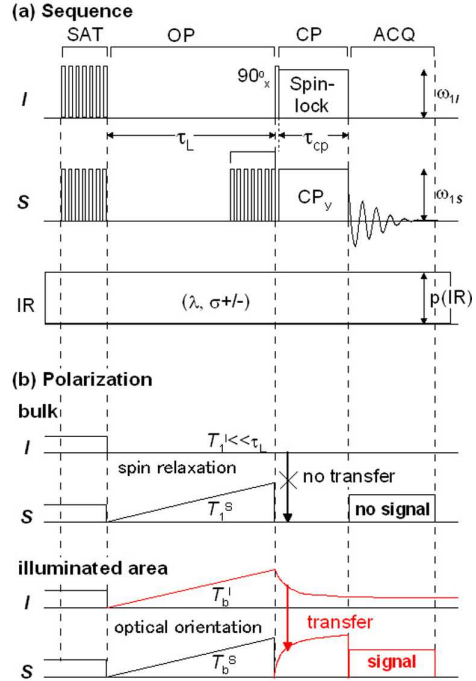}
}
 \end{center}
\caption{
(a) Pulse sequence for the \textit{surface-sensitive NMR}. 
$I$ and $S$: primary and secondary nuclei. IR: infrared light
with the wavelength $\lambda$ and the helicity $\sigma^{\pm}$.
(b) Time dependences of the $I$ and $S$ polarizations
in the bulk and in the illuminate area (schematic). 
Only the $S$ signal in the illuminated area is detected.
Note that the horizontal axes are not in scale.
}
\label{sequence}
\end{figure}

In some cases, 
it is possible to identify signals from a part of a bulk material 
even in the absence of field gradients,
by making use of the structural features.
For example, at interfaces in composite materials,
nuclear polarization can be transferred from one material to another.
In this case, 
the transferred nuclear polarization is localized in proximity to the interfaces 
within the nuclear spin diffusion length, 
so that one can selectively detect NMR signals from that area.
This can be referred to as \textit{interface-sensitive NMR}.
The method is often combined with dynamic nuclear polarization or optical pumping.
Examples include solid surfaces adsorbed by hyperpolarized $^{129}$Xe gas
\cite{driehuys93,bowers93,long93}, 
solvents in dynamically polarized solutions\cite{kusaka07} 
and interfaces between solid materials and semiconductors.
\cite{tomaselli99,goehring03}

Our current interest resides in the last case,
especially that combined with the optical pumping.\cite{tycko98}
We proposed a \textit{nuclear spin polarizer} (NSP)
as a possible configuration of this kind,\cite{goto03a,goto04a} 
which utilize the successive polarization transfer scheme
shown in Fig. \ref{spt}.
It is intended to enhance signals from a surface of a (target) material 
to which conventional dynamic nuclear polarization techniques are inapplicable.
With the aid of an optically oriented semiconductor, 
nuclei in the target material are indirectly polarized.
Examples of the target include nano-structures 
built on semiconducting bulk materials,
such as quantum dots on substrates.
A typical configuration of the scheme is depicted in Fig. \ref{illumination}.

\section{\textit{Surface-sensitive NMR}}
\label{procedure}

Here, we introduce a possible variation of the \textit{nuclear spin polarizer},
i.e., \textit{surface-sensitive NMR}
in optically pumped semiconductors.
Target nuclei ($S$) are located 
at a surface of a bulk compound semiconductor,
whose signals are detected apart from those in the rest of the material.
A key feature of the scheme is that 
it utilizes the primary nuclei ($I$)
with a long spin-lattice relaxation time $T_1^I$
(as is often the case in semiconductors)
as a polarization reservoir for the secondary (target) nuclei ($S$) to be detected,
which enables us to detect $S$ signals from the illuminated area selectively,
as illustrated in Fig. \ref{ss-nmr}.
This is in contrast to the conventional (bulk) NMR in a uniform magnetic field, 
where nuclei in the whole sample are excited and detected.

The pulse sequence of the scheme is shown in Fig. \ref{sequence}
together with the time dependence of the $I$ and $S$ polarizations 
in both the illuminated area and the rest of the sample (bulk).
The sequence consists of the four processes.
(1) Saturation (SAT): 
comb pulses made of trains of 90$^{\circ}$-pulses
are applied to both the nuclei 
to extinguish the initial polarizations in thermal equilibrium.
(2) Optical pumping (OP): 
the infrared light is irradiated for the effective duration time of $\tau_{\rm L}$
to create polarizations in the illuminated area.
In this period, the $S$ polarization in the bulk 
may recover toward its thermal equilibrium value with the time scale of $T_1^{S}$,
while that of $I$ does not because $T_1^{I}$ is much longer than $\tau_{L}$.
(3) Cross polarization (CP): 
the second SAT pulse is applied for $S$, which is followed by
a cross polarization between $I$ and $S$.
The CP promotes polarization transfers in the illuminated area,
but not in the bulk because the $I$ polarization is absent there.
(4) Signal acquisition (ACQ): 
only the $S$ signal in the illuminated area is detected 
in the final signal acquisition process. 

\section{\textit{Surface-sensitive NMR} in InP}
\label{results}
We have previously reported successive polarization transfer experiments 
in indium phosphide.\cite{goto06a,goto08a}
Here, we discuss the results 
in the context of the \textit{surface-sensitive NMR}.
The sample used was a 350 $\mu$m-thick wafer of the semi-insulating iron-doped InP
with the crystal orientation of (100) 
and the carrier density at room temperature $\rho = 7 \times 10^7$ cm$^{-3}$.
It was set with the surface normal to the magnetic field
and a light beam.

\begin{figure}[b]
\begin{center}
\resizebox{0.4\textwidth}{!}{%
 \includegraphics{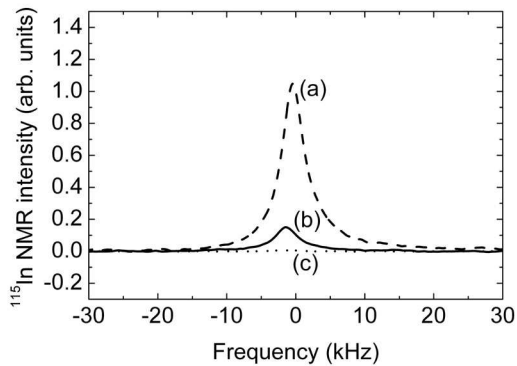}
}
\end{center}
\caption{
$^{115}$In NMR spectra in iron-doped InP at $T=$10 K and $H_0=$6.346 T.
The spectrum (a) was obtained with the sequence 
SAT$-\tau_L-(90^{\circ})_x-$FID without light irradiation,
which represents a bulk signal in thermal equilibrium.
The spectra (b) and (c) were obtained
with the sequence shown in Fig. \ref{sequence},
with and without light irradiation ($\lambda=$890 nm and $\sigma^{+}$), respectively.
The spectrum (b) represents a signal in \textit{surface-sensitive NMR},
and the absence of signals in (c) demonstrates that the origin of 
the signal in (b) is purely the optically oriented nuclei.
Resonance conditions were as follows:
$\omega_{0I}/2\pi$ = 109.316 MHz, $\omega_{0S}/2\pi$ = 59.23 MHz,
$\omega_{1I}/2\pi = \omega_{1S}/2\pi$ = 18 kHz.
}
\label{spectra}
\end{figure}

Figure \ref{spectra} shows three Fourier-Transform spectra 
of $^{115}$In at 6.346 T and 10 K. 
The spectrum (a) was obtained with the sequence 
SAT$-\tau_L-(90^{\circ})_x-$FID without IR irradiation, 
which represents a bulk $^{115}$In signal in thermal equilibrium.
The spectrum (b), on the other hand, 
was obtained with the sequence shown in Fig. \ref{sequence}
with $I=^{31}$P, $S=^{115}$In and $\tau_{\rm L}$=120 s.
The wavelength ($\lambda$) and helicity ($\sigma^{\pm}$) 
were set at 890 nm and $\sigma^+$, respectively,
where the largest optical orientation effects were observed 
in both $^{31}$P and $^{115}$In.
Since the photon energy is below the band gap,
the penetration depth of the light may be as thick as a few tens of $\mu$m.
\cite{michal99}

The relaxation ($T_1$) and the buildup ($T_\textnormal{b}$) times 
were reported to be
$T_1^S \approx 0.1$ s, $T_1^I \approx 10^5$ s and 
$T_\textnormal{b}^S \approx T_\textnormal{b}^I \approx 10^3$ s,\cite{goto08a,goto04b}
hence the condition $T_1^I \gg T_\textnormal{b}^I$ is satisfied.
As a result, transfer from the bulk $I$ spins is absent
and the spectrum (b) represents only the signal from the illuminated area, 
as explained in Fig. \ref{sequence} (b).

The spectrum (c) was obtained with the same sequence as (b) 
but without light irradiation.
Since no polarization grows during $\tau_L$, 
no signal intensity is observed in (c).

The total efficiency of the scheme, 
which is evaluated by the polarization of the target nuclei,
is maximized by optimizing each of the polarization transfer processes.
In the cross polarization process,
the transferred polarization is generally expressed as,\cite{mehring83} 
\begin{equation}
 M_S(\tau_{cp})\propto 
[ 1-\exp \{-(1-\frac{T_{IS}}{T_{1\rho}^{I}})\frac{\tau_{cp}}{T_{IS}}\} ]
\exp(-\frac{\tau_{cp}}{T_{1\rho}^{I}}),
\label{MS}
\end{equation}
where $T_{IS}$ and $T_{1\rho}$ 
are the cross-relaxation time between $I$ and $S$
and the relaxation time of $I$ in the rotating flame, respectively.
As is evident from (\ref{MS}), 
the cross-relaxation time ($T_{IS}$) 
should be faster than the relaxation time ($T_{1\rho}$)
to minimize transfer loss during the cross polarization process.

The present material (iron-doped InP) is fortunate in this respect.
First, the relaxation in the rotation flame is negligible 
($T_{1\rho}^{I} \rightarrow \infty$)
due mainly to the high crystal symmetry.\cite{tomaselli98}
Secondly, the cross relaxation time is rather fast under the optical pumping condition.
It was reported to be $T_{IS}=(1.1 \pm 0.1) \times 10^{-4}$ s,\cite{goto08a}
which is by an order of magnitude faster than that estimated for the nuclear dipolar couplings,
because of the indirect $J$-couplings of the order of kHz.\cite{goto08a}.
These result indicates that semi-insulating iron-doped InP
is a good candidate for the \textit{nuclear spin polarizer}.

\section{Conclusion}
\label{conclusion}
In this article, we have described the scheme of the \textit{surface-sensitive NMR} 
in optically pumped semiconductors
and shown its feasibility with the example of InP wafers.\cite{goto06a,goto08a}
The scheme provides a basis for the \textit{nuclear spin polarizer},
an interface-sensitive NMR in optically pumped semiconductors.
\cite{tycko98,goto03a,goto04a}

\section*{Acknowledgments}
This work was supported at its early stage
by Industrial Technology Research Grant Program in 2002 
from New Energy and Industrial Technology Development Organization (NEDO) of Japan.
A.G. acknowledges support by Grant-in-Aid for Basic Research 
from Japan Society for Promotion of Science (JSPS).
S.O. acknowledges support from the Nanotechnology Support Project 
of the Ministry of Education, Culture, Sports, Science and Technology (MEXT) of Japan. 
T.S. appreciates support 
from the International Center for Materials Nanoarchitectonics, NIMS, Japan.

%

\end{document}